\documentclass{SCIS2019}
\usepackage{epstopdf}
\usepackage{color}
\usepackage{graphicx}

\begin{document}
%\oa
%%%%%%%%%%%%%%%%%%%%%%%%%%%%%%%%%%%%%%%%%%%%%%%%%%%%%%%
%%% Authors do not modify the information below
%%% 作者不需要修改此处信息
\ArticleType{LETTER}
%\SpecialTopic{}
\Year{2019}
\Month{}
\Vol{}
\No{}
\DOI{}
\ArtNo{}
\ReceiveDate{}
\ReviseDate{}
\AcceptDate{}
\OnlineDate{}
%%%%%%%%%%%%%%%%%%%%%%%%%%%%%%%%%%%%%%%%%%%%%%%%%%%%%%%

%%% title: 标题
%%%   \title{title}{title for citation}
\title{Secure Analysis Over Generalized-$K$ Channels}{Secure analysis over generalized-$K$ channels}

%%% Corresponding author: 通信作者
%%%   \author[number]{Full name}{{email@xxx.com}}
%%% General author: 一般作者
%%%   \author[number]{Full name}{}
\author[1]{Luyao ZHANG}{}
\author[2]{Hui ZHAO}{}
\author[3]{Gaofeng PAN}{}
\author[4]{Liang YANG}{{liangyang.guangzhou@gmail.com}}
\author[1]{Jiawei CHEN}{}
%\author[3]{Ddd AUTHOR}{}

%%% Author information for page head. 页眉中的作者信息
\AuthorMark{Zhang L}

%%% Authors for citation. 首页引用中的作者信息
\AuthorCitation{Zhang L, Zhao H, Pan G, et al}

%%% Authors' contribution. 同等贡献
%\contributions{Authors A and B have the same contribution to this work.}

%%% Address. 地址
%%%   \address[number]{Affiliation, City {\rm Postcode}, Country}
\address[1]{School of Mathematics and Statistics, Southwest University, Chongqing {\rm 400715}, China}
\address[2]{Computer, Electrical, and Mathematical Science and Engineering  Division,\\ King Abdullah
University of Science and Technology, Thuwal {\rm 23955-6900}, Saudi Arabia}
\address[3]{School of Information and Electronics, Beijing Institute of Technology, Beijing {\rm 100081}, China}
\address[4]{College of Computer Science and Electronic Engineering, Hunan University, Changsha {\rm 410082}, China}

\maketitle

%%%%%%%%%%%%%%%%%%%%%%%%%%%%%%%%%%%%%%%%%%%%%%%%%%%%%%%
%%% The main text. 正文部分
%%%%%%%%%%%%%%%%%%%%%%%%%%%%%%%%%%%%%%%%%%%%%%%%%%%%%%%
\begin{multicols}{2}
\deareditor
The secrecy outage probability (SOP) is used to estimate the secrecy outage performance when the transmitter has no state information about potential wiretap channels \cite{Bloch}, i.e., silent eavesdropping. Most previous work, such as~\cite{Lei_CL,Wang_CL}, has focused on the SOP defined in \cite{Bloch}, which we call the conventional SOP in this letter, where both unreliable transmission from the transmitter to the legitimate receiver (i.e., outage) and information leakage to eavesdroppers are considered to be secrecy outage. Thus, according to the conventional SOP definition, a secrecy outage does not necessarily imply that any information has been leaked to eavesdroppers. To capture the actual information leakage, \cite{Zhou} proposed a new SOP definition, which we call the proposed SOP in this letter, that is exactly the information leakage probability. However, they only considered the SOP over Rayleigh fading channels, a simple small-scale channel model. In actual wireless communications scenarios, shadowing is typically involved, resulting in large-scale fading \cite{Zhao_CL}.

The generalized-$K$ (GK) fading model was proposed \cite{Bithas} to capture composite fading channels (with both small- and large-scale fading), but the exact model involves modified Bessel functions of the second kind, which usually results in Meijer's G-function or more advanced special functions appearing in the final SOP expression \cite{Lei_TVT}. It is still a matter of debate as to whether the Meijer's G-function can be viewed as a closed-form expression. To avoid this function, \cite{Atapattu} simplified the GK model by using a mixed Gamma distribution.

Although the SOP over GK fading channels has already been investigated \cite{Lei_CL,Lei_TVT}, the authors in \cite{Lei_CL,Lei_TVT} only considered the conventional SOP definition, which does not give the actual information leakage probability. Moreover, they did not investigate the asymptotic performance when the main link's signal-to-noise ratio (SNR) is sufficiently large, which gives the secrecy diversity order and array gain \cite{Wang_CL}. Others have studied the SOP's asymptotic behavior \cite{Lei_FITEE}, but their conclusions as to its secrecy diversity order are not valid in the general case.

In this letter, we adopt the SOP definition in \cite{Zhou} and the simplified model of \cite{Atapattu}, and derive a closed-form expression for the proposed SOP over GK fading channels. To simplify this expression and obtain additional insights, we also perform an asymptotic analysis of the main link in the high-SNR region.

\lettersection{System Model}
In the standard Wyner model \cite{Bloch}, a source transmits confidential messages to a destination $d$. Meanwhile, an eavesdropper $e$ wants to overhear this information. Here, we assume that all links undergo independent GK fading. The exact probability density function (PDF) $\gamma_t$ of the instantaneous SNR at $t$ ($t \in \{d, e\}$) is given by \cite{Lei_TVT}
\begin{align}
{f_{{\gamma _t}}}\left( x \right) = \frac{{G_{0,2}^{2,0}\left( {\frac{{{k_t}{m_t}x}}{{{{\overline \gamma  }_t}}}\left| {_{{k_t},{m_t}}^ - } \right.} \right)}}{{\Gamma \left( {{k_t}} \right)\Gamma \left( {{m_t}} \right)}x},
\end{align}
where $k_t$ and $m_t$ are the parameters of the GK fading channels and $\overline \gamma_t$ is the average of $\gamma_t$. In addition, $G(\cdot)$ and $\Gamma(\cdot)$ represent the Meijer's G-function and the Gamma function, respectively.

To avoid the appearance of Meijer's G-function in the final SOP expression, we adopt the simplified model of \cite{Atapattu}, where the PDF and cumulative density function (CDF) of $\gamma_t$ are \cite{Atapattu,Lei_CL}
\begin{align}
{f_{{\gamma _t}}}\left( x \right) &= \sum\limits_{{j_t} = 1}^L {{a_{t,{j_t}}}} {x^{{m_t} - 1}}\exp \left( { - {\zeta _{t,{j_t}}}x} \right),\\
{F_{{\gamma _t}}}\left( x \right) &= 1 - \sum\limits_{{j_t} = 1}^L {\sum\limits_{{n_t} = 0}^{{m_t} - 1} {\frac{{{A_{t,{j_t}}}{{\left( {{\zeta _{t,{j_t}}}x} \right)}^{{n_t}}}\exp \left( { - {\zeta _{t,{j_t}}}x} \right)}}{{{n_t}!}}} },
\end{align}
respectively, where ${a_{t,{j_t}}} = \frac{{{\theta _{t,{j_t}}}}}{{\sum\nolimits_{v = 1}^L {{\theta _{t,v}}\Gamma \left( {{m_t}} \right)\zeta _{t,v}^{ - {m_t}}} }}$, ${\theta _{t,{j_t}}} = \frac{{{k_t}{m_t}{\omega _{{j_t}}}t_{{j_t}}^{{k_t} - {m_t} - 1}}}{{{t_{{j_t}}}{{\overline \gamma  }_t}\Gamma \left( {{m_t}} \right)\Gamma \left( {{k_t}} \right)}}$, ${\zeta _{t,{j_t}}} = \frac{{{k_t}{m_t}}}{{{t_{{j_t}}}{{\overline \gamma  }_t}}}$, and ${A_{t,{j_t}}} = \Gamma \left( {{m_t}} \right)\sum\nolimits_{v = 1}^L {{a_{t,v}}} \zeta _{t,v}^{ - {m_t}}$. In addition, $L$, $\omega_{j_t}$ and $t_{j_t}$ are the number of terms in the sum, weight factors, and abscissas for the Gauss--Laguerre integration, respectively.

\lettersection{Secrecy Outage Probability}
We consider the silent eavesdropping case, where the source does not have access to the channel state information about the wiretap channel. In this case, perfect security cannot be guaranteed, due to the constant confidential information rate $R_s$ in the source's encoder.

The proposed SOP \cite{Zhou} is given by
\begin{align}
{\rm{SOP}} &= \Pr \left\{ {{C_e} > {C_d} - {R_s}\left| {{\gamma _d} > \mu } \right.} \right\} \notag\\
%& = \Pr \left\{ {{\gamma _d} < \lambda  - 1 + \lambda {\gamma _e}\left| {{\gamma _d} > \mu } \right.} \right\} \notag\\
& = \frac{{\Pr \left\{ {\mu  < {\gamma _d} < \lambda  - 1 + \lambda {\gamma _e}} \right\}}}{{\Pr \left\{ {{\gamma _d} > \mu } \right\}}},
\end{align}
where $\lambda=2^{R_s}$ and ${C_t} = {\log _2}\left( {1 + {\gamma _t}} \right)$ ($t \in \{d,e\}$) denotes the capacity of $t$'s channel. In this definition, $\mu$ is chosen so as to achieve reliable transmission from the source to the destination, striking a compromise between the quality of service at the destination and communication security. If $\mu=0$, this SOP reduces to the conventional SOP \cite{Bloch}.

The proposed SOP can be rewritten as
\begin{align}\label{SOP_int}
%&{\rm{SOP}} = \frac{{\int_{\frac{{\mu  + 1}}{\lambda } - 1}^\infty  {\int_\mu ^{\lambda  - 1 + \lambda {x}} {{f_{{\gamma _d}}}\left( y \right)} dy} {f_{{\gamma _e}}}\left( x \right)dx}}{{{{\overline F }_{{\gamma _d}}}\left( \mu  \right)}} \notag\\
%& = \frac{{\int_{\frac{{\mu  + 1}}{\lambda } - 1}^\infty  {\left[ {{F_{{\gamma _d}}}\left( {\lambda  - 1 + \lambda {\gamma _e}} \right) - {F_{{\gamma _d}}}\left( \mu  \right)} \right]} {f_{{\gamma _e}}}\left( x \right)dx}}{{{{\overline F }_{{\gamma _d}}}\left( \mu  \right)}} \notag\\
{\rm{SOP}}=& \frac{{\int_{\frac{{\mu  + 1}}{\lambda } - 1}^\infty  {{F_{{\gamma _d}}}\left( {\lambda  - 1 + \lambda {x}} \right)} {f_{{\gamma _e}}}\left( x \right)dx}}{{{{\overline F }_{{\gamma _d}}}\left( \mu  \right)}} \notag\\
&- \frac{{{F_{{\gamma _d}}}\left( \mu  \right){{\overline F }_{{\gamma _e}}}\left( {\frac{{\mu  + 1}}{\lambda } - 1} \right)}}{{{{\overline F }_{{\gamma _d}}}\left( \mu  \right)}},
\end{align}
where $\overline F_{\gamma_t}(\cdot)$ represents the complementary CDF (CCDF) of $\gamma_t$, which is equal to $1-F_{\gamma_t}(\cdot)$.

After a certain amount of mathematical manipulation, the proposed SOP over GK fading channels can be rewritten as
\begin{align}
&{\rm{SOP}} = {{\overline F}_{{\gamma _e}}}\left( {\frac{{\mu  + 1}}{\lambda } - 1} \right) - \frac{1}{{{{\overline F}_{{\gamma _d}}}\left( \mu  \right)}}\sum\limits_{{j_d} = 1}^L {\sum\limits_{{n_d} = 0}^{{m_d} - 1} {\sum\limits_{{j_e} = 1}^L {\frac{1}{{{n_d}!}}} } } \notag\\
&{A_{d,{j_d}}}{a_{e,{j_e}}}\zeta _{d,{j_d}}^{{n_d}}\exp \left( { - {\zeta _{d,{j_d}}}\left( {\lambda  - 1} \right)} \right)\sum\limits_{f = 0}^{{n_d}} {n_d \choose f} {\lambda ^f} \notag\\
&\frac{{{{\left( {\lambda  - 1} \right)}^{{n_d} - f}}\Gamma \left( {{m_e} + f,\left( {{\zeta _{e,{j_e}}} + {\zeta _{d,{j_d}}}\lambda } \right)\left( {\frac{{\mu  + 1}}{\lambda } - 1} \right)} \right)}}{{{{\left( {{\zeta _{e,{j_e}}} + {\zeta _{d,{j_d}}}\lambda } \right)}^{{m_e} + f}}}},
\end{align}
where $\Gamma(\cdot,\cdot)$ denotes the upper incomplete Gamma function.

\lettersection{Asymptotic Analysis}
By letting $\overline \gamma_d$ go to infinity while $\overline \gamma_e$ remains finite, we can derive the asymptotic SOP (ASOP), which governs the SOP's behavior at high SNRs and gives the secrecy diversity order and array gain \cite{Wang_CL}.

When $\overline \gamma_d \to \infty$ and $m_d \neq k_d$, the asymptotic CDF of $\gamma_d$ can be written as \cite{Wang_CL}
\begin{align}
F_{{\gamma _d}}^\infty \left( x \right) = \frac{{\Gamma \left( {\left| {{k_d} - {m_d}} \right|} \right)}}{{\Gamma \left( {{k_d}} \right)\Gamma \left( {{m_d}} \right)v}}{\left( {\frac{{{k_d}{m_d}x}}{{{{\overline \gamma  }_d}}}} \right)^v} + o\left( {\overline \gamma  _d^{ - v - 1}} \right),
\end{align}
where $o(\cdot)$ denotes the higher order terms and $v = \min \left\{ {{k_d},{m_d}} \right\}$. (Due to space limitations, we do not consider the asymptotic CDF for $m_d=k_d$, where the diversity order is also $v$.)

Setting $\overline F_{\gamma_d}(\cdot)=1$ in the SOP definition \eqref{SOP_int} yields
\begin{align}
{\rm{SO}}{{\rm{P}}^\infty } =& \int_{\frac{{\mu  + 1}}{\lambda } - 1}^\infty  {F_{{\gamma _d}}^\infty \left( {\lambda  - 1 + \lambda x} \right)} {f_{{\gamma _e}}}\left( x \right)dx \notag\\
&- F_{{\gamma _d}}^\infty \left( \mu  \right){\overline F _{{\gamma _e}}}\left( {\frac{{\mu  + 1}}{\lambda } - 1} \right).
\end{align}
This integral-form expression for the ASOP is valid for general fading channels and can be regarded as complementing the work of \cite{Zhou}, who did not investigate the asymptotic behavior.

After certain mathematical manipulations, we can derive the ASOP as
\begin{align}
{\rm{SO}}{{\rm{P}}^\infty } =& \frac{{\Gamma \left( {\left| {{k_d} - {m_d}} \right|} \right){{\left( {{k_d}{m_d}} \right)}^v}}}{{\Gamma \left( {{k_d}} \right)\Gamma \left( {{m_d}} \right)v}}\left[ {\sum\limits_{f = 0}^v {v \choose f} {{\left( {\lambda  - 1} \right)}^{v-f}}} \right. \notag\\
&{\lambda ^f}\sum\limits_{{j_e} = 1}^L {\frac{{{a_{e,{j_e}}}\Gamma \left( {{m_e} + f,{\zeta _{e,{j_e}}}\left( {\frac{{\mu  + 1}}{\lambda } - 1} \right)} \right)}}{{\zeta _{e,{j_e}}^{{m_e} + f}}}} \notag\\
&\left. { - {\mu ^v}{{\overline F}_{{\gamma _e}}}\left( {\frac{{\mu  + 1}}{\lambda } - 1} \right)} \right]\overline \gamma _d^{ - v},
\end{align}
from which we can see that the secrecy diversity order is $v$.

\lettersection{Numerical Results}
\begin{figure*}
\centering
\includegraphics[width= 6 in]{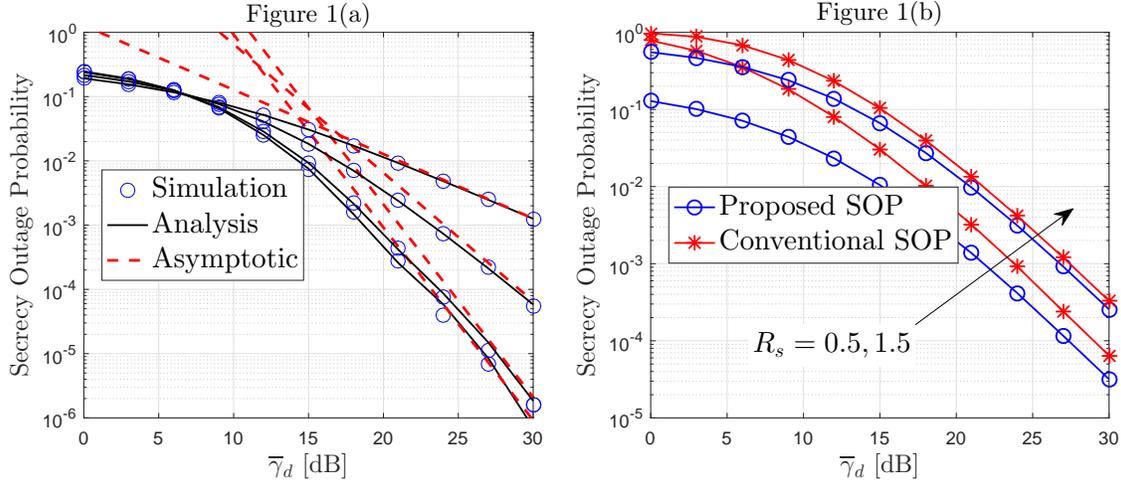}
\caption{SOP versus $\overline \gamma_d$ for $\overline \gamma_e=0$ dB, $k_d=k_e=3$, $R_s=1$, $\mu=3$ and $L=15$ in (a), and $\overline \gamma_e=1$ dB, $k_d=k_e=3$, $m_d=m_e=2$ and $\mu=3$ in (b).}
\end{figure*}
In this section, we conduct Monte Carlo simulations to validate the correctness of our derived expressions. In Figure 1(a), we can see that the SOP improves as $\overline \gamma _d$ increases, due to the better state of the main channel. It decreases as $m_d=m_e$ grows, due to the increasing number of paths over the main channel. In addition, the SOP's slope clearly changes for $m_d=m_e=1,2$ but remains constant for $m_d=m_e=4,5$, showing that the secrecy diversity order is $\min\{k_d,m_d\}$.

Figure 1(b) shows the probability gap between the proposed and conventional SOP definitions. Although the trends are similar, the conventional SOP cannot provide the exact probability of leaking information to eavesdroppers. It is also worth noting that the probability gap converges as $R_s$ increases.

\lettersection{Conclusion}
In this letter, we have derived a closed-form expression for the SOP defined in \cite{Zhou} over GK fading channels, as well as the corresponding asymptotic result, valid for high SNRs, which gives the secrecy diversity order and array gain. We also provide a general integral form for the ASOP, which can be used to derive the asymptotic result for any given fading channel. Finally, we have presented numerical results that demonstrate the accuracy of our derived expressions and show the probability gap between the two SOP definitions.

%%%%%%%%%%%%%%%%%%%%%%%%%%%%%%%%%%%%%%%%%%%%%%%%%%%%%%%
%%% Acknowledgements. 致谢
%%%%%%%%%%%%%%%%%%%%%%%%%%%%%%%%%%%%%%%%%%%%%%%%%%%%%%%
\Acknowledgements{
This work was in part supported by the National Natural Science Foundation of China (NSFC) under Grant 61671160, the Department of Education of Guangdong Province (No. 2016KZDXM050), the Hunan Natural Science Foundation under Grant (No. 2019JJ40043), the Fundamental Research Funds for the Central Universities, and the open research of the State Key Laboratory of Advanced Optical Communication Systems and Networks, Shanghai Jiao Tong University, China (No. 2019GZKF03004)}

%%%%%%%%%%%%%%%%%%%%%%%%%%%%%%%%%%%%%%%%%%%%%%%%%%%%%%%
%%% Supplements. 补充材料, 非必选
%%%%%%%%%%%%%%%%%%%%%%%%%%%%%%%%%%%%%%%%%%%%%%%%%%%%%%%
%\Supplements{Appendix A.}

%%%%%%%%%%%%%%%%%%%%%%%%%%%%%%%%%%%%%%%%%%%%%%%%%%%%%%%
%%% Reference section. 参考文献
%%% citation in the content using "some words~\cite{1,2}".
%%% ~ is needed to make the reference number is on the same line with the word before it.
%%% Please make sure there are no more than 9 items of references.
%%%%%%%%%%%%%%%%%%%%%%%%%%%%%%%%%%%%%%%%%%%%%%%%%%%%%%%

\end{multicols}
\end{document}